\documentclass[a4paper]{jpconf}
\usepackage{graphicx}
\usepackage{verbatim}
\usepackage{algorithm}
\usepackage{listings}
\usepackage{array}

\begin{document}
\title{CutLang: a cut-based HEP analysis description language and runtime interpreter}

\author{G. Unel$^1$, S. Sekmen$^2$, A.M. Toon$^3$,}

\address{$^1$ University of California at Irvine, Department of Physics and Astronomy, Irvine, USA}
\address{$^2$ Kyungpook National University, Department of Physics, Daegu, South Korea}
\address{$^3$ Saint Joseph University of Beirut, Computer Software Engineering, Beirut, Lebanon}

\ead{gokhan.unel@cern.ch}

\begin{abstract}
We present CutLang, an analysis description language and runtime interpreter for high energy collider physics data analyses.  
An analysis description language is a declerative domain specific language that can express all elements of a data analysis in an easy and unambiguous way.  A full-fledged human readable analysis description language, incorporating logical and mathematical expressions, would eliminate many programming difficulties and errors, consequently allowing the scientist to focus on the goal, but not on the tool.  In this paper, we discuss the guiding principles and scope of the CutLang language, implementation of the CutLang runtime interpreter and the CutLang framework, and demonstrate an example of top pair reconstruction.

\end{abstract}

\section{Introduction}

Nowadays, any physicist performing analysis with the Large Hadron Collider (LHC) data needs to be well-versed in programming, at the level of both a system programmer and a software developer.  Processing vast amounts of collision and simulation events is a tedious task where simplest programming mistakes can create big confusions on the analysis results. Moreover, a multitude of different analysis frameworks for similar tasks make it difficult to communicate analysis algorithms  within the same experiment and to preserve them for discussions outside the experiment. The steep learning curve for analysis frameworks also leads to very long lead-times for analyses, erecting a barrier between data and the physicist who may simply wish to try an analysis idea.  All these difficulties could be overcome by creating analysis description languages (ADLs), which are human readable, declerative domain specific languages capable of describing the analysis flow in a standard and unambiguous way, independent of any computing framework.  

An ADL would decouple the mathematical and logical algorithm of a physics analysis from computing operations, and being declerative, would eliminate programming difficulties and errors, consequently allowing the analysis to be more efficient and flawless.  Working with a unique domain specific language rather than a general purpose language like c++ or python has the advantage of introducing standardization and unambiguity in expressing analysis components.  This would make comparing and combining different analyses within an experiment, or between experiments significantly easier.  An ADL would also assist phenomenologists in understanding and reinterpreting analyses, as well as easily suggesting new analysis ideas for testing physics models.  Moreover, an ADL can be used for educating high school students or other enthusiasts about the structure of a collider analysis. It would also serve the long term preservation of analyses, by providing standardized and accessible description of the analyses for the whole high energy physics (HEP) community.  

In summary, an ADL would facilitate the abstraction, design, visualization, validation, combination, reproduction, interpretation and overall communication of HEP analyses. If widely accepted for use in the HEP community, an ADL would clearly maximize the scientific output of the LHC or any upcoming long term international experiment like the Future Circular Collider.  Today, the abundance of inexpensive and easily accessible computing power is leading to a fundamental paradigm shift in data analysis.  The development of a full-fledged domain specific ADL combined with the increasing accessibility of computing power could eventually evolve into an AI-based analysis tool with which scientists can easily communicate in their own languages.

In this note, we present CutLang, a complete ADL, and a runtime interpreter for easily writing and performing collider  analyses~\cite{Sekmen:2018ehb}. A simple top quark pair reconstruction analysis is also shown as an example to demonstrate the implementation of the CutLang ADL and to illustrate the advantages of the ADL approach.

\section {CutLang, the language}
CutLang's development precedes community wide discussions and synthesis efforts such as~\cite{Brooijmans:2016vro}. The ADL in CutLang was designed to have a number of properties aiming to serve the needs of experimental, phenomenology and education communities in HEP, such as human readability, correctness, easiness in learning, and programming language and framework independence. CutLang ADL describes the physics algorithm of event processing in an analysis in a plain text file using syntax rules that include standard mathematical and logical operations and 4-vector algebra. CutLang syntax was designed to be brief, easily readable and understandable. The syntax follows a keyword-value structure. The physics scope consists of analysis object definitions, observable definitions and event selections, which are organized in dedicated blocks. The following summarize the core physics content that can be defined within the CutLang ADL.
\begin{description}
\item [Predefined analysis objects:] Types for basic input particles are predefined in CutLang ADL.
\item [Simple analysis objects:] An arbitrary number of simple objects like jets, electrons, muons can be defined based on cuts on object attributes such as transverse momentum, pseudorapidity, heavy flavor tagging discriminators, etc.  
\item [Derived analysis objects:] Objects can be defined from already existing objects (e.g. defining b-tagged jets from high transverse momentum jets).
\item [Composite analysis objects:] Composite objects can be reconstructed from multiple simple objects (e.g. reconstruct a Z boson from two electrons).
\item [Event selection regions:] Arbitrary number of event selection regions can be defined based on event selection criteria.  Multiple selection regions can be defined from already defined regions (e.g. defining signal regions and control regions from a preselection region).  
\item [Concurrency in event selection:] Concurrent definition of multiple algorithms on the same set of events is possible.
\item [Simple constants or event variables:] Keywords can be assigned to constants (e.g. Z boson mass) or variables (e.g. angular variables between objects, mass of the Z boson reconstructed from two leptons, etc.).  
\item [Complex variables via external functions:] An analysis can contain complex variables that cannot be expressed using simple algebraic operations, or that could be numerical.  In CutLang, such variables are defined in self-contained external C++ functions, which can be referenced from the ADL text file.  
\item [Histogramming:] Since CutLang is also a runtime interpreter, it also allows to define histograms for event variables in the ADL file at any stage of event selection.  1- and 2-dimensional histograms can be specified and filled at run time using ROOT libraries.  Histograms are defined with the keyword \texttt{histo}.
\end{description}
The following mathematical and logical operations are available for defining objects and event selection regions:
\begin{description}
\item [Comparison operators and thresholds:] All standard comparison operators, i.e. $==, <, >, =<, >=$ as well as two additional interval operators for inclusion ($[\,]$) and exclusion ($]\,[$), both defined by International standard ISO 31-11~\cite{iso}, 
are available for comparing particle or event properties to limiting values.   
\item [Logical operators:] Both symbolic and alphanumerical forms of the AND(\&\&) and OR ($||$) operators. 
\item [Ternary operator:] Application of conditional selection criteria is available, including nested statements. The C++ syntax is assumed: {\it condition $?$ true-case $:$ false-case }.  
\item [$\chi^2$ minimization] In an event, particles best fulfilling a specified criterion can be selected using an optimization algorithm. The indices of the particles that would be determined at run time are to be specified as negative integers.
\item [Simple mathematical operators:] The following operations can be used in variable definitions, or in selection statements: $+$, $-$, $\ast$, $/$, \^{}, sin(), cos(), tan(), abs(), sqrt().
\end{description}
Usage of the language elements and the syntax will be illustrated by a top quark pair reconstruction analysis example in Section~\ref{sec:ttrecoexample}.

\section{CutLang, the runtime interpreter}

From the start, CutLang was designed to contain a runtime interpreter to bypass the inherent inefficiency of the modify-compile-run cycle.  In an interpreted analysis system, adding new selection criteria, changing the execution order or cancelling analysis steps are more practical.  CutLang runtime interpreter is written in C++ based on ROOT classes for handling Lorentz vector operations, input file and histogram manipulations, and operates in any modern Unix-like environment. The interpreter is compiled with ROOT libraries only once in the beginning or when optional external functions for complex variables are added. The actual parsing of the ADL text relies on automatically generated dictionary and grammar based on traditional tools Lex and Yacc~\cite{lexandyacc}. The ADL file is split into tokens by Lex and the hierarchical structure of the algorithm is found by Yacc. 

\section{CutLang framework and tools}

The CutLang framework includes the CutLang interpreter and additional tools and facilities for performing a typical LHC analysis.  The framework can read event data from multiple input formats such as ATLAS and CMS Open Data, DELPHES, FCC, etc. given as ROOT files. New input data types can be integrated using the tools and prescriptions provided inside the framework. All event types are converted into predefined input particle types, where the particles are pre-sorted in the order of decreasing transverse momentum, and the most energetic particle is the zeroth one. The missing transverse energy ($E_T^{miss}$) in the event is also mapped to a Lorentz vector with zero axial momentum.  Additionally, neutrinos coming from boosted W boson decays are assigned a special object, which is a massless and chargeless particle with transverse momentum and azimuthal angle ($\phi$) values extracted from $E_T^{miss}$, but its pseudorapidity is assumed equal to that of the associated charged lepton.

CutLang framework also includes many predefined functions that are standard in an analysis, such as the invariant mass of particles, or angular distance measures between particles.  The complete list can be found in~\cite{Sekmen:2018ehb}.  Moreover, external user functions for calculating complex variables can be incorporated into the framework.

CutLang provides output in ROOT format, with all algorithm results and analysis histograms in separate directories in the resulting files.  These directories also contain user definitions, derived objects and analysis cut-flows as text output, and event selection efficiencies as a histogram.  Output of surviving events in ntuples format is planned for the upcoming release.


\begin{table}
\centering
\caption{$t\bar{t}$ reconstruction user definitions}
{\small
\begin{tabular}{|c|>{}p{6.8cm}|l|} \hline 
Step & Explanation & Definition syntax in CutLang \tabularnewline \hline  \hline 
1 & WH1 is reconstructed from 2 jets & \texttt{\footnotesize{}def WH1 : JET\_-1 JET\_-1}\tabularnewline \hline 
2 & WH2 is reconstructed from 2 other jets  & \texttt{\footnotesize{}def WH2 : JET\_-11 JET\_-11}\tabularnewline \hline 
3 & mWH1 is the mass of WH1  & \texttt{\footnotesize{}def mWH1 : \{WH1 \}m}\tabularnewline \hline 
4 & mWH2 is the mass of WH2  & \texttt{\footnotesize{}def mWH2 : \{WH2 \}m}\tabularnewline \hline 
5 & mTopH1 is the mass of  WH1 and another jet & \texttt{\footnotesize{}def mTopH1 : \{WH1 JET\_-2 \}m }\tabularnewline \hline 
6 & mTopH2 is the mass of WH2 and another jet& \texttt{\footnotesize{}def mTopH2 : \{WH2 JET\_-4 \}m }\tabularnewline \hline 
7 & WHbR1 is the angular separation between WH1 and its associated jet & \texttt{\footnotesize{}def WHbR1 : \{WH1 , JET\_-2 \}dR}\tabularnewline \hline 
8 & WHbR2 is the angular separation between WH2 and its associated jet & \texttt{\footnotesize{}def WHbR2 : \{WH2 , JET\_-4 \}dR}\tabularnewline \hline 
9 & topchi2 is the two top quark candidates mass difference squared  & \texttt{\footnotesize{}def topchi2 :((mTopH1 - mTopH2)/4.2)\textasciicircum 2}\tabularnewline \hline 
10 & Wchi2 is the squared sum of the deviation of W boson candidates from known value & \texttt{\footnotesize{}def Wchi2 :((mWH1-80.4)/2.1)\textasciicircum 2+((mWH2-80.4)/2.1)\textasciicircum 2 }\tabularnewline \hline 
\end{tabular}
}
\label{tab:userdefs}
\end{table}

\section{An example physics analysis \label{sec:ttrecoexample}}
Top quark pair reconstruction is a frequent procedure in both standard model and new physics analyses at the LHC.  
A typical implementation for top pair reconstruction in a 6 jet final state, with no explicit use of b-jet tagging, performed in two alternative methods, would consist of the following steps:

\begin{itemize}
\item Define user keywords, shortcuts to be used in the analysis, e.g. for hadronic W bosons and top quarks, their masses,  angular distances between them, and $\chi^2$ variables (see Table~\ref{tab:userdefs}).
\item Define the preselection region, consisting of 6 jets and low $E_T^{miss}$ requirements common to both algorithms (see Table~\ref{tab:preselection}).
\item Perform the single-step search algorithm, which finds the correct jet combinations for top pair reconstruction in a single step; impose further W mass and angular variable cuts; and plot some variables (see Table~\ref{tab:singlestep}).
\item Perform the alternative double-step search algorithm, which finds the correct jet combinations for top pair reconstruction in two steps, by first reconstructing the W bosons, and then finding the best associated b jet candidates; impose further W mass and angular variable cuts; and plot some variables (see Table~\ref{tab:doublestep}).
\end{itemize}

\begin{table}
\centering
\caption{ The preselection algorithm}
{\small
\begin{tabular}{|c|>{}p{7.1cm}|l|} \hline 
Step & Explanation & Commands in CutLang \tabularnewline \hline  \hline 
1 & name this algorithm "preselection"  & \texttt{\footnotesize{}algo preselection} \tabularnewline \hline 
2 &  count all events & \texttt{\footnotesize{}cmd ALL  }\tabularnewline \hline 
3 &  select events with 6 or more jets & \texttt{\footnotesize{}cmd nJET >= 6 }\tabularnewline \hline 
4 &  select events with $E_T*{miss}$ less than 100 GeV& \texttt{\footnotesize{}cmd MET < 100 }\tabularnewline \hline 
\end{tabular}
}
\label{tab:preselection}
\end{table}

\begin{table}
\centering
\caption{ The "singlestep" algorithm}
{\small
\begin{tabular}{|c|>{}p{6.5cm}|l|} \hline 
Step & Explanation & Commands in CutLang \tabularnewline \hline  \hline 
1 & name this algorithm "singlestep"  & \texttt{\footnotesize{}algo singlestep } \tabularnewline \hline 
2 & execute preselection algorithm & \texttt{\footnotesize{}preselection } \tabularnewline \hline 
3 & minimize the total $\chi^2$ and find all 6 jets & \texttt{\footnotesize{}cmd topchi2 + Wchi2 $\sim$= 0} \tabularnewline \hline 
4 &mWH1 should be between 50 and 120 GeV &\texttt{\footnotesize{}cmd mWH1 [] 50 120 } \tabularnewline \hline 
5 &mWH2 should be between 50 and 120 GeV &\texttt{\footnotesize{}cmd mWH2 [] 50 120 } \tabularnewline \hline 
6 &WHbR1 should be greater than 0.6 &\texttt{\footnotesize{}cmd WHbR1 > 0.6 } \tabularnewline \hline 
7 &WHbR2 should be greater than 0.6  &\texttt{\footnotesize{}cmd WHbR2 > 0.6 } \tabularnewline \hline 
8 &Histogram the mass of WH1 &\texttt{\footnotesize{}histo "mWHh1, W2 mass (GeV), 50, 50, 150, mWH1" } \tabularnewline \hline 
9 &Histogram the mass of WH2 &\texttt{\footnotesize{}histo "mWHh2, W1 mass (GeV), 50, 50, 150, mWH2"  } \tabularnewline \hline 
10 &Histogram the mass of TopH1 &\texttt{\footnotesize{}histo "mTopHh1, top1 mass (GeV), 70, 0, 700, mTopH1" } \tabularnewline \hline 
11 & Histogram the mass of TopH2&\texttt{\footnotesize{}histo "mTopHh2, top2 mass (GeV), 70, 0, 700, mTopH2" } \tabularnewline \hline 
\end{tabular}
}
\label{tab:singlestep}
\end{table}

\begin{table}
\centering
\caption{ The "doublestep" algorithm}
{\small
\begin{tabular}{|c|>{}p{6.5cm}|l|} \hline 
Step & Explanation & Commands in CutLang \tabularnewline \hline  \hline 
1 &name this algorithm "doublestep"&\texttt{\footnotesize{}algo doublestep  } \tabularnewline \hline 
2&execute preselection algorithm&\texttt{\footnotesize{}preselection } \tabularnewline \hline 
3&minimize the W$\chi^2$ & \texttt{\footnotesize{}cmd Wchi2 $\sim$= 0  } \tabularnewline \hline 
4&minimize the top $\chi^2$ &\texttt{\footnotesize{}cmd topchi2 $\sim$= 0  } \tabularnewline \hline 
5&mWH1 should be between 50 and 120 GeV&\texttt{\footnotesize{}cmd mWH1 [] 50 120  } \tabularnewline \hline 
6&mWH2 should be between 50 and 120 GeV&\texttt{\footnotesize{}cmd mWH2 [] 50 120  } \tabularnewline \hline 
7&WHbR1 should be greater than 0.6 &\texttt{\footnotesize{}cmd WHbR1 > 0.6 } \tabularnewline \hline 
8&WHbR2 should be greater than 0.6  &\texttt{\footnotesize{}cmd WHbR2 > 0.6 } \tabularnewline \hline 
  9&Histogram the mass of WH1&\texttt{\footnotesize{}histo "mWHh1, W1 mass (GeV), 50, 50, 150, mWH1" } \tabularnewline \hline 
10&Histogram the mass of WH2&\texttt{\footnotesize{}histo "mWHh2, W2 mass (GeV), 50, 50, 150, mWH2"  } \tabularnewline \hline 
11&Histogram the mass of TopH1&\texttt{\footnotesize{}histo "mTopHh1, top1 mass (GeV), 70, 0, 700, mTopH1" } \tabularnewline \hline 
12&Histogram the mass of TopH2&\texttt{\footnotesize{}histo "mTopHh2, top2 mass (GeV), 70, 0, 700, mTopH2"} \tabularnewline \hline 
\end{tabular}
}
\label{tab:doublestep}
\end{table}

\section{Conclusions}
Increases in both the number and complexity of analysis tasks in collider experiments, as well as advancements in computing methods and resources are leading to new analysis approaches.  One such approach is the utilization of domain-specific analysis description languages to define all analysis elements in an easy and unambiguous way.  CutLang, the first implementation of such a domain specific ADL capable of expressing the complete set of event processing operations in a collider physics data analysis as a runtime interpreted language continues to be developed with addition of new features.  CutLang aims to serve as a regular tool for the high energy community in general: from experimental analysts and phenomenologists to educators in areas from analysis design to preservation. The ultimate goal is to develop CutLang into an AI-based analysis tool with which scientists can easily communicate in their own languages.

\end{document}